\documentclass[a4paper,12pt,twoside,fleqn]{article}
\usepackage{epsfig,times,fancyheadings,isolatin1}
\usepackage[longnamesfirst,sort]{natbib} % Better bibliography.
\usepackage[normalem]{ulem} % Better underlining (but use italics for
\bibpunct{(}{)}{;}{a}{}{,}
\normalsize % Set \baselineskip
\newenvironment{tcenter}{\begin{list}{}{\setlength{\leftmargin}{22mm}%
      \setlength{\rightmargin}{22mm}\setlength{\topsep}{0pt}%
      \setlength{\parskip}{0pt}\setlength{\partopsep}{0pt}%
      }\item[]\centering}{\end{list}}

\makeatletter
\renewcommand{\section}{%
  \@startsection{section}{1}{0pt}{\baselineskip}{1pt}{\centering}}
\renewcommand{\subsection}{%
  \@startsection{subsection}{2}{0pt}{0pt}{-1ex}{\uline}}
\renewcommand{\@seccntformat}[1]{{\csname the#1\endcsname}.~}
\makeatother
\renewcommand{\sectionmark}[1]{}
\renewcommand{\subsectionmark}[1]{}

\newcommand{\captionfonts}{\small}
\makeatletter  % Allow the use of @ in command names
\long\def\@makecaption#1#2{%
  \vskip\abovecaptionskip
  \sbox\@tempboxa{{\captionfonts #1: #2}}%
  \ifdim \wd\@tempboxa >\hsize
    {\captionfonts #1: #2\par}
  \else
    \hbox to\hsize{\hfil\box\@tempboxa\hfil}%
  \fi
  \vskip\belowcaptionskip}
\makeatother   % Cancel effect of \makeatletter

\newenvironment{Figure}[1][hbtp]{\begin{figure}[#1]\begin{list}{}{%
        \setlength{\leftmargin}{5mm}\setlength{\rightmargin}{5mm}%
        }\item[]}{\end{list}\vspace{-\baselineskip}\end{figure}}

\begin{document}
\thispagestyle{plain}
%\vspace*{3.4cm}
\vspace*{52mm}\vspace*{-\baselineskip}
\begin{tcenter}
%ATOMIC-SCALE SIMULATIONS OF\\ NANOCRYSTALLINE METALS\\[\baselineskip]
  SIMULATIONS OF NANOCRYSTALLINE METALS AT THE ATOMIC SCALE.  WHAT CAN
  WE DO?  WHAT CAN WE TRUST?\\[\baselineskip]
  J. Schi{\o}tz\\[\baselineskip] 
  Center for Atomic-scale Materials Physics and\\ Department of Physics,
  Technical University of Denmark,\\ DK-2800 Lyngby, Denmark.
\end{tcenter}\vspace{\baselineskip}

\section*{ABSTRACT}

In recent years it has become possible to study the properties of
nanocrystalline metals through atomic-scale simulations of systems
with realistic grain sizes.  A brief overview of the main results is
given, such as the observation of a reverse Hall-Petch effect --- a
softening of the metal when the grain size is reduced.  The
limitations of computer simulations are discussed, with a particular
focus on the factors that may influence the reliability of this kind
of simulations.

\section{INTRODUCTION}

Molecular dynamics and other atomic-scale simulation techniques are
providing much new understanding of various phenomena in materials
science.  A significant effort has been made to model nanocrystalline
metals using these techniques.  The purpose and goals of such
modelling is at least twofold.  On one hand, the materials themselves
have technologically interesting properties, and a better
understanding of nanocrystalline materials is hoped to have direct
technological relevance.  On the other hand, these materials provide a
simplified model system useful for testing and developing theories
about deformation in polycrystalline materials in general.  The
exceedingly small grain size offers two advantages
for theoretical studies.  One is a suppression of large-scale
structures in the grains, hopefully leading to a ``simpler'' material,
the other is the possibility for direct atomic-scale simulation.

Significant progress has been made in our understanding of the
structure and mechanics of nanocrystalline metals.  For example, the
grain boundaries are found to participate directly in the deformation
process of nanocrystalline metals, and at sufficiently small grain
sizes they appear to be carrying the majority of the deformation.

Less progress has been made on generalizing the results of
atomic-scale simulations from nanocrystalline metals to more coarse
grained metals.  Direct extrapolation is likely to give a
misleading picture, as new processes become active as the grain size
is increased from the nanoscale.  Many of the processes dominating the
mechanics of coarse grained metals cannot occur within the tiny
grains of a nanocrystalline metal.

This paper reviews the progress that has been made in the last few
years using computer simulations to model nanocrystalline metals,
and discusses the reliability of such simulations.  For a more general
review of nanocrystalline metals, see for example \citet{Mo98}.

\section{ATOMIC-SCALE SIMULATION TECHNIQUES}

The main atomics-scale simulation techniques that have been applied to
nanocrystalline metals are molecular dynamics and energy minimization
techniques.  In both cases, one starts with a description of the
potential energy of the system expressed in the form of a
``potential'', i.e.\ a function yielding the potential energy of an
atom as a function of the positions of its neighbours.  The potential
must be carefully chosen in order to provide a realistic description
of the bonding in the metal.

In molecular dynamics, one solves Newton's second law numerically for
the atoms in the system.  Energy is thus preserved, and one obtains
the positions (and velocities) of all the atoms as a function of time
--- the challenge is then to extract useful information from these
data.  Molecular dynamics can easily be modified to provide a
well-defined temperature instead of a constant total energy.

Energy minimization techniques, sometimes known as ``molecular
statics'' follow a similar idea.  Instead of solving Newton's second
law the potential energy of the system is minimized with respect to
all atomic coordinates.  If this is done while changing an external
parameter (for example the shape of the system) one obtains a
``zero-temperature'' simulation, describing how the system evolves in
response to the external influence at a temperature of zero Kelvin.

The results of an atomic-scale computer simulation is typically many
megabytes of atomic coordinates, velocities, and possibly other
information such as atomic-scale stresses.  Computerized analysis and
visualisation tools are clearly necessary to make sense of such
amounts of data.  Sometimes the data is analyzed by calculating
quantities similar to those obtained from experiments (for example
radial distribution functions), and then use well-established
experimental techniques to extract information about the structure of
the simulated material.  This is particularly useful if the calculated
quantity is to be compared directly with experiments; but in many
cases it goes against the ``spirit'' of atomic-scale simulations: that
atomic-scale information is already present, and should be used
directly.

Plotting the atomic structure directly, by drawing the atoms in a
slice of the sample, is very useful, but must usually be augmented
with further data analysis, such as selecting certain classes of atoms
and/or coloring atoms according to stress, energy, displacement during
the simulation, etc.  A technique that has proven particularly useful
in the context of nanocrystalline metals is the Common Neighbor
Analysis (CNA), where the local crystalline structure around each atom
is identified \citep{FaJo94,HoAn87}.  This makes it relatively easy to
identify grain boundaries, dislocation cores, stacking faults, etc.

\section[]{MAIN RESULTS FROM ATOMIC-SCALE SIMULATIONS\nopagebreak\\ OF
  NANOCRYSTALLINE METALS}

A number of papers reporting atomic-scale simulations of nanophase
materials have been published in recent years \citep[see for
example][]{Ch95, PhWoGl95, PhWoGl95b, ZhAv96, KePhWoGl97, SwCa97,
  SwCa98, SwSpCaFa99, SwFaCa00, DeSw01, ScDiJa98, ScVeDiJa98,
  ScVeDiJa99, HeRi01}.  In this paragraph some of the main results are
reviewed.

\subsection{Mechanical properties of nanocrystalline fcc metals.}
\label{sec:mechanical}

The mechanical properties of nanocrystalline copper, palladium and
nickel have mainly been studied by Van Swygenhoven et al.\ and by
Schiøtz et al.  The three metals behave in a very similar way.

A series of simulations of plastic deformation of nanocrystalline
copper has been published by \citet{ScVeDiJa99}.  The structure of one
of the simulated systems is shown in Fig.~\ref{fig:showsys}.  The
structure shows evidence of some dislocation activity, such as the
extrinsic stacking fault in Fig.~\ref{fig:showsys}b, which has been
left behind by two partial dislocations (Shockley partials) moving
through the grain.  A detailed analysis of the dislocation activity
during the deformation process indicates that the dislocations cannot
be responsible for more than a quarter of the observed plastic
deformation.  The remainder of the plastic deformation is caused by a
large number of apparently uncorrelated sliding events in the grain
boundaries, see \citet{ScVeDiJa99} for details.
\begin{Figure}
  \begin{center}
    \epsfig{file=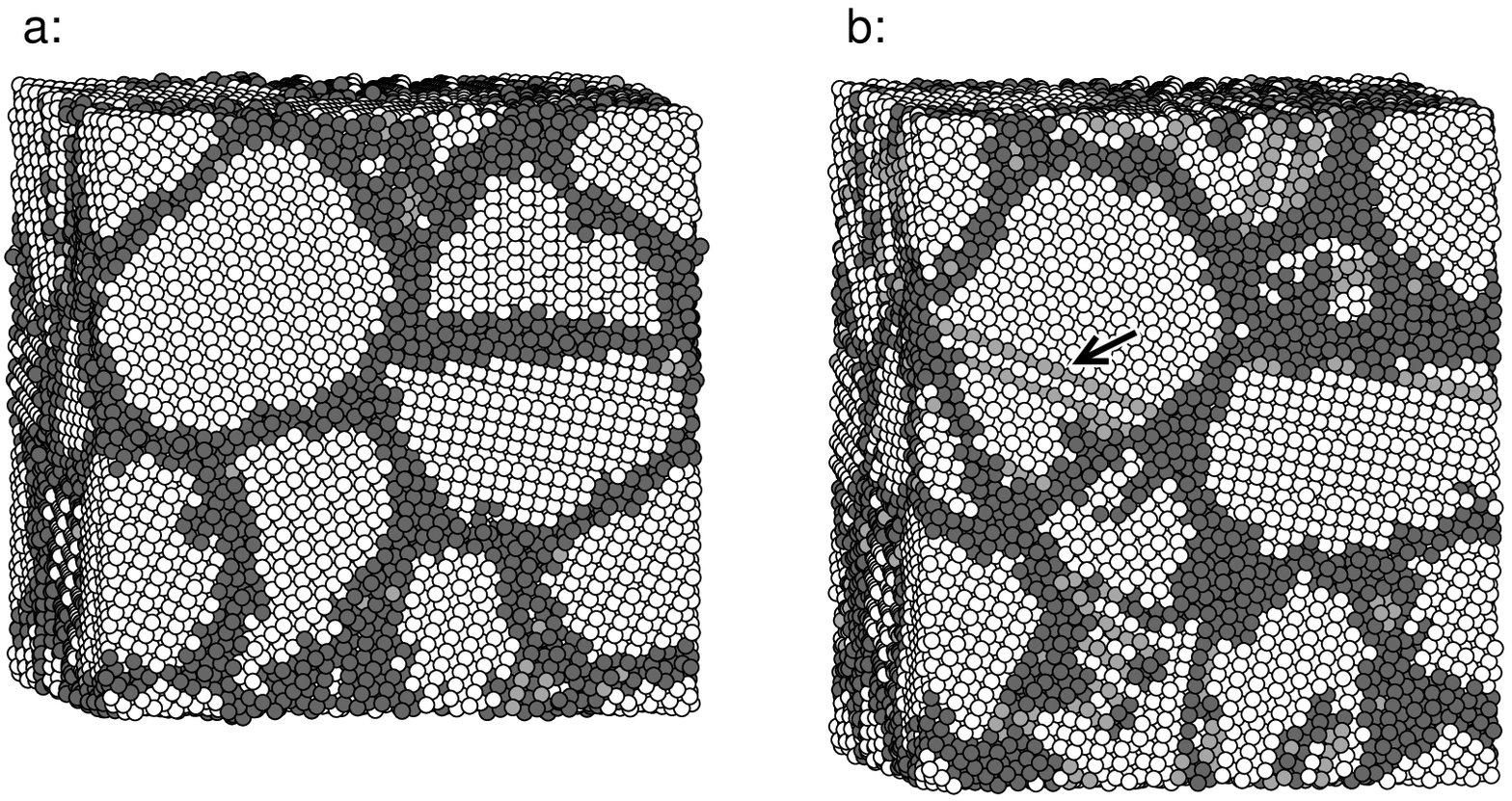, width=0.8\linewidth}
    \caption{The initial (a) and final (b) configuration of a
      nanocrystalline copper system deformed 10\% at 300\,K.  The
      system contains approximately 100\,000 atoms arranged in 16
      grains, giving an average grain size of 5.2\,nm.  Systems with
      smaller grain sizes were simulated with approximately the same
      number of atoms, while systems with larger grain sizes were
      simulated with ten times as many atoms to keep the number of
      grains sufficiently large.  The arrow indicates where two
      partial dislocations have moved through a grain, leaving an
      extrinsic stacking fault behind.  The majority of the plastic
      deformation has occurred in the grain boundaries.}
    \label{fig:showsys}
  \end{center}
\end{Figure}
\begin{Figure}
  \begin{center}
    \raisebox{\depth}{%
      \epsfig{file=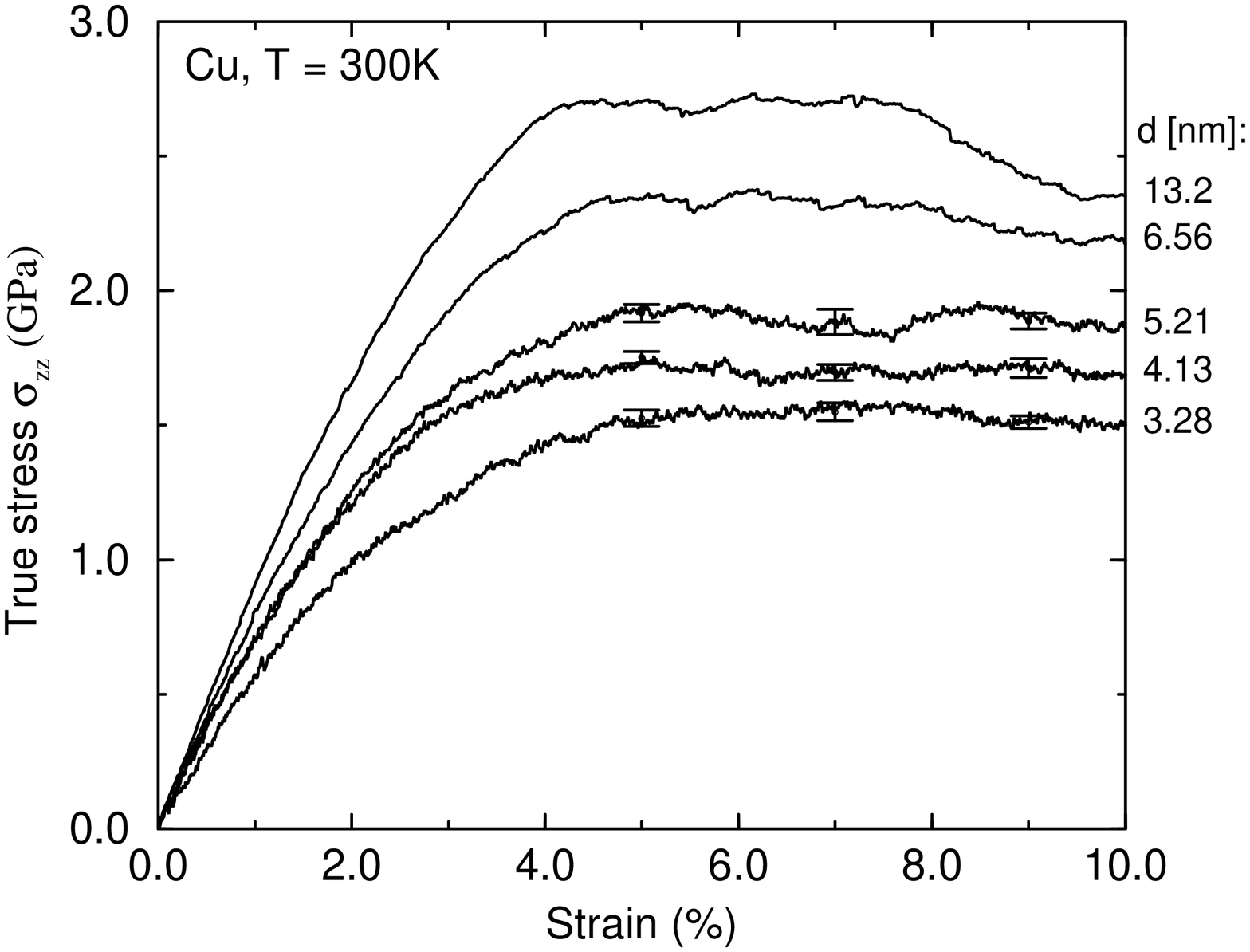, angle=-90, totalheight=7.5cm}}
    \hfill
    \raisebox{\depth}{%
      \epsfig{file=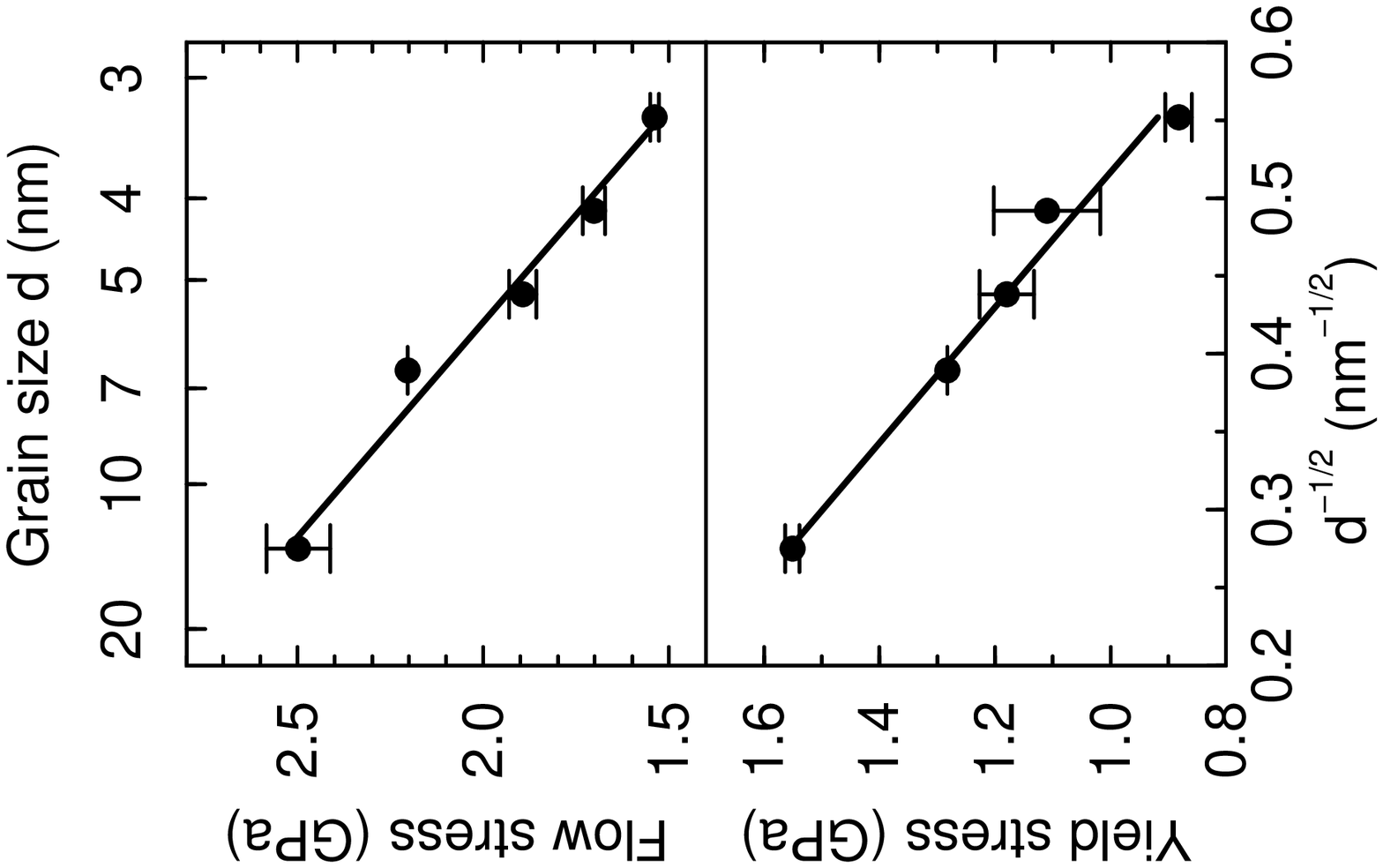, angle=-90, totalheight=8.5cm}}
    \caption{Stress-strain curves for nanocrystalline copper at 300K
      for varying grain sizes.  The yield and flow stress is seen to
      decrease with decreasing grain size, this is summarized to the right.
      Adapted from \citet{ScVeDiJa99}.}
    \label{fig:stressCu}
  \end{center}
\end{Figure}

The simulations resulted in realistic stress-strain curves, see
Fig.~\ref{fig:stressCu}a.  A systematic variation of the mechanical
properties with grain size is seen in Fig.~\ref{fig:stressCu}a, and
summarized in Fig.~\ref{fig:stressCu}b.  The metal becomes harder
as the grain size is increased.  This is called the \emph{reverse}
Hall-Petch effect, since the opposite behavior is usually seen in
metals with larger grain sizes \citep{Ha51,Pe53}.

In coarse-grained metals, the plastic deformation is carried by
dislocations, and the grain boundaries act as barriers to the
dislocations.  Reducing the grain size thus reduces the mobility of
the dislocations, and leads to a harder metal.  In nanocrystalline
metals with sufficiently small grain sizes, the major part of the
plastic deformation is carried by the grain boundaries --- at least
according to the computer simulations.  It is, therefore, not
surprising that decreasing the grain size leads to a softening of the
metal, as the volume fraction of the grain boundaries increases.
Based on the simulations, one can therefore predict that there should
be an ``optimal'' grain size, where the hardness of a metal is
maximal, see Fig.~\ref{fig:optimal}.
\begin{Figure}
  \begin{center}
    \begin{minipage}[t]{0.4\linewidth}
      \epsfig{file=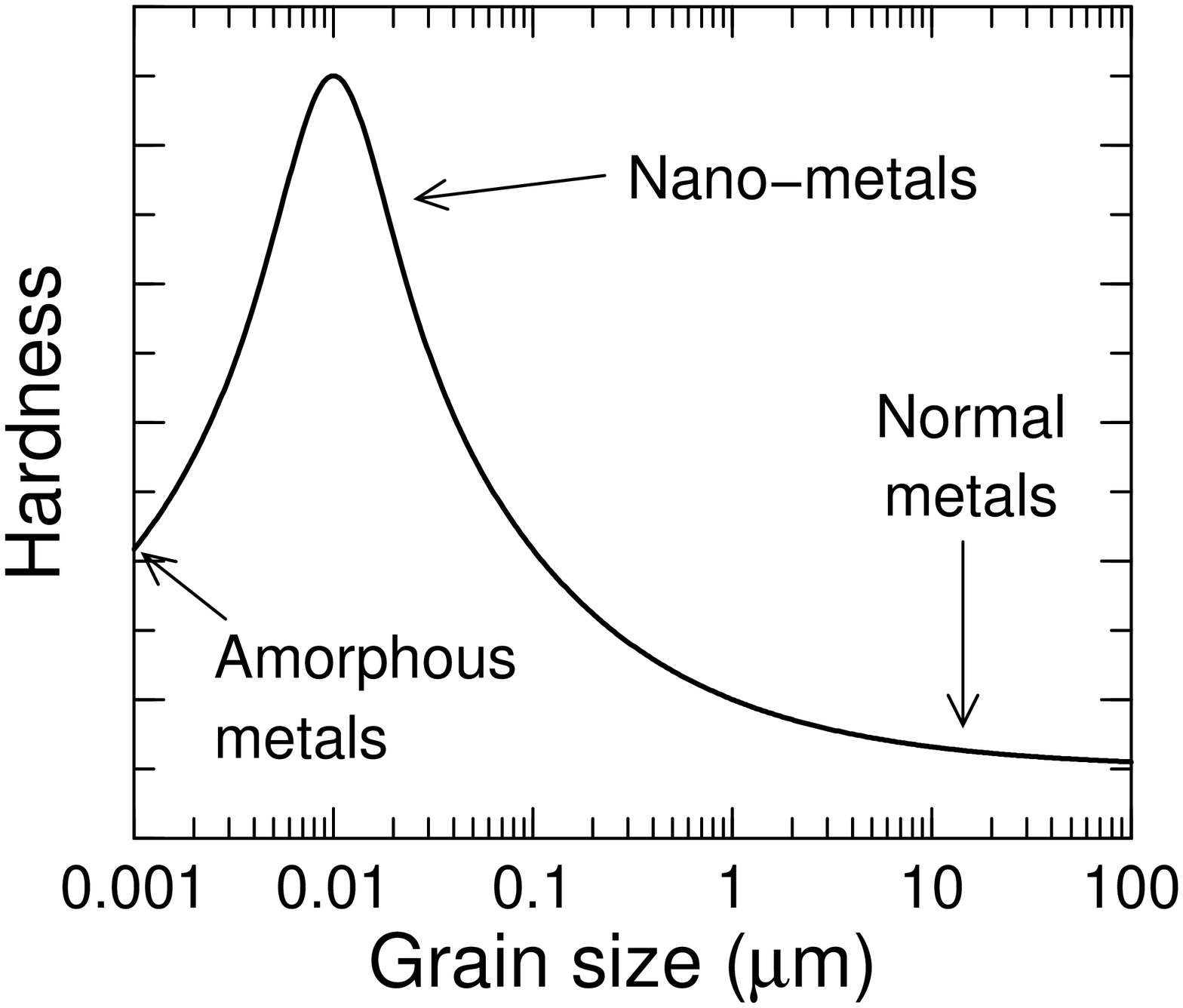, bb=104 59 588 628, angle=-90, width=\linewidth}
    \end{minipage}\hfill
    \begin{minipage}[t]{0.55\linewidth}
      \caption{Hardness of a metal as a function of the grain size.
        In the large-grain limit the hardness is given by the
        Hall-Petch relation $H = H_\infty + k d^{-1/2}$, in the other
        limit the hardness increases with grain size, indicating the
        existence of a maximum in hardness.  On the figure the maximum
        has rather arbitrarily been placed at 10\,nm.  The position of
        the maximum is expected to vary with the material and probably
        also with strain rate, but is expected to be in the tens of
        nanometers.  See also \citet{NiWa91} and \citet{Yi98}.}
      \label{fig:optimal}
    \end{minipage}
  \end{center}
\end{Figure}

The simulations of \citet{ScDiJa98,ScVeDiJa99} were performed at
constant strain rate, typically $5 \times 10^8 {\rm s}^{-1}$ (see
section \ref{sec:scales} for a discussion of these high strain rates).
In the simulation one dimension of the simulation cell (the length)
was controlled, while the two lateral dimensions were allowed to
evolve in order to keep the lateral components of the stress near
zero.  The stress was calculated during the simulation to obtain the
stress-strain curves.  Van Swygenhoven et al.\ 
(\citeyear{SwCa97,SwCa98,SwSpCaFa99}) have performed simulations of
plastic deformation of nanocrystalline nickel, where the applied
\emph{stress} has been the control parameter: a fixed stress was
applied, and the dimensions of the simulation cell was allowed to
evolve in time, yielding a ``measured'' strain rate.  Strain rates in
the order of $10^8 {\rm s}^{-1}$ were seen; decreasing with increasing
grain size.  The observed grain size dependence of the strain rate is
evidence of a reverse Hall-Petch effect.

\subsection{Possible evidence for multiple deformation mechanisms.}
\label{sec:multiple}

Assuming that the stress-strain relationship in nanocrystalline nickel
is similar to the relationship in copper (see Fig.~\ref{fig:stressCu})
one would expect a stress-dependent critical grain size for plastic
deformation.  If, for example, nanocrystalline copper were loaded with
a tensile stress of 2\,Gpa, one would expect that samples with a grain
size below $\approx$6\,nm would flow, whereas samples with a larger
grain size will exhibit some initial deformation which quickly stops.
However, this is not quite what is observed by Van Swygenhoven et al.
The strain rate is seen to decrease with increasing grain size, but at
larger grain sizes it becomes approximately constant at a value of
$\approx 10^7 {\rm s}^{-1}$, independent of grain size.  This is taken
as evidence for a change in deformation mechanism \citep{SwSpCaFa99}.
Apparently, a slower deformation mechanism is active in the regions of
the stress-strain diagram in Fig.~\ref{fig:stressCu} which appear to
be elastic.  Evidence for such a mechanism was also reported by
\citet{ScVeDiJa99}:  If the deformation is stopped while still in the
``elastic'' region, the stress is seen to relax exponentially with a
relaxation time near 100\,ps.

There is thus some evidence for two different, but closely related
deformation mechanisms.  The atomic-scale mechanisms appear to be very
similar in the two cases, the main part of the deformation is carried
by the grain boundaries \citep{SwSpCaFa99,ScVeDiJa99}.  In the faster
of the two mechanisms some dislocation activity is also seen in the
grains.  It is not clear if the slow mechanism can give rise to large
deformations, or if it stops after a small strain has been created.
It is possible that the slow deformation mechanism is not able to
accommodate the small changes in grain shape, so the grains cannot
efficiently slide past each others, whereas the limited dislocation
activity seen in the faster mechanism allows such shape changes.  This
is, however, pure speculation as no computer simulations have been
made where large deformations are obtained at a strain rate near $10^7
{\rm s}^{-1}$ --- molecular dynamics simulations of large systems is
limited to times below a few nanoseconds.  It is not unlikely that the
two deformation mechanisms are essentially the same, but that similar
small deformation events in the grain boundaries can be triggered both
by thermal fluctuations and by the high stress.  The two deformation
mechanisms then corresponds to thermal activation or stress activation
being dominant.

Unfortunately, there is no simulational evidence for the transition
from the reverse to the normal Hall-Petch regime.  It is expected that
above a critical grain size, the majority of the deformation will be
carried by dislocations moving inside the grains.  However, little
dislocation activity is seen even for the largest grain sizes.  It
is likely that an increase of the grain size from two to five times
will be necessary to observe the normal Hall-Petch effect, requiring
10--100 times as many atoms, and correspondingly larger computers.

\subsection{The structure of nanocrystalline metals.}
\label{sec:structure}

The structure of the grain boundaries in nanocrystalline metals has
been a subject generating significant debate.  It has been proposed
that grain boundary structures of nanocrystalline metals are
radically different from those in coarse grained metals.  For
example, a non-equilibrium low density structure has been proposed
\citep{ZhBiHeGl87}, but later experimental work is indicating that the
grain boundaries are of high density, and not too different from those
in coarse grained metals \citep{FiEaMuWa91,StSiNeSaHa95}.
Nevertheless, there are significant uncertainties concerning the
detailed grain boundary structure.  A number of simulations have been
addressing this question
\citep{PhWoGl95,PhWoGl95b,SwFaCa00,ZhAv96,KeWoPhGl99}, reaching
various conclusions.  Recently, \citet{KeWoPhGl99} found that the
grain boundary structure is glassy without much structure.  They
obtain this result by analysing the radial pair-distribution function
of the atoms in the grain boundaries.  \Citet{SwFaCa00} find that the
grain boundaries contain a lot of structure, with segments looking like
coincidence site lattice boundaries, separated by more disordered
regions.  They base their conclusions on direct visualisation of the
grain boundary structure, and argue that the less direct method of
analysing the pair-distribution function is likely to depend on the
algorithm chosen to select the grain boundary atoms.

The question of the exact structure of the grain boundaries does not
appear to have been settled yet, but it should not be forgotten that
the structure found in a simulation may depend on the algorithm used
to set up the grain structure in the first place (see section
\ref{sec:initialconf}).  If the structure turns out to be independent
of grain size, these results will also be of relevance for general
grain boundaries in coarse-grained polycrystalline metals.

\section{FROM THE ATOMIC SCALE TO THE MACRO-SCALE.}
\label{sec:macro}

Atomic-scale simulations cannot directly be applied to the deformation
of everyday metals, as the length scale of many important processes is
far larger than anything that can be simulated at the atomic scale.
Simulating a volume of $1 \mu{\rm m}^3$ for a single second would
require computers that are $10^{12}$ times larger than anyone
available today.  Nevertheless, atomic-scale simulations can
contribute to the understanding of these materials as well.

Materials processes that are important in both nanocrystalline and
coarse-grained metals can be observed in simulations of the
former.   One example could be emission of dislocations from grain
boundaries and triple junctions, which is occasionally seen in the
simulations.   A detailed study of those processes might reveal useful
information about them, but no such work has to my knowledge been
published.

A more promising approach is to directly model fundamental processes
that are known to be important, and where atomic-scale simulations may
provide new information.  One example is the modelling of cross-slip
of screw dislocations and annihilation of screw dislocation dipoles
\citep{RaJaLePeSrJo97,VeRaLePeJa00,VeRaLePeJa01}.  Further discussion
of this and related work is beyond the scope of this paper.

\section{THE RELIABILITY OF ATOMIC-SCALE SIMULATIONS.}
\label{sec:reliability}

The main advantage of atomic-scale computer simulations is the
unrestricted access to all atomic coordinates.  In principle,
everything that happens can be determined with confidence from the
motion of all the atoms --- although in reality the amount of
information is so large that computerized analysis and visualisation
is necessary to extract information from the simulations.

The main disadvantage of computer simulations, atomic-scale or
otherwise, is that one cannot be sure how well they represent the
reality they are supposed to reflect.  This section discusses the
main questions of reliability that must be considered in connection
with atomic-scale simulations of nanocrystalline metals.

\subsection{The precision and completeness of the underlying model.}

All computer simulations are based on an underlying set of assumptions
and equations.  Clearly, if the underlying model is incorrect, one
cannot expect the computer simulation to give useful results.  Many
computer simulations rely on a large set of (explicit or implicit)
assumptions about the processes going on in the system under
investigation.  For example, simulations in materials science often
make assumption about the way plastic deformation occurs, how grains
grow, etc; the weather forecast assumes a set of equations describing
the hydrodynamics and thermodynamics of the atmosphere. In many cases
the assumptions are well-founded and well-tested, but there is always
the possibility that they do not apply in a given situation.  For
example, a simulation assuming that all plastic deformation occurs
through the motion of dislocations would not have been able to find
the reverse Hall-Petch effect mentioned previously.

One of the main virtues of molecular dynamics and related atomic-scale
simulation methods is that the set of assumptions is very small.  The
simulation method does not assume anything about the processes
occurring inside the material, as it is based on numerically
integrating Newton's second law while observing the motion of the
atoms.  The simulation will therefore not be biased by the
user's more or less well-founded assumptions about the
processes occurring inside the material.  For example, none of the
simulations presented in section \ref{sec:mechanical} assume that
grain boundaries can slide or that dislocations can more (or even that
dislocations exist, they appear ``by themselves'').  Even relatively
exotic phenomena such as phonon-induced friction on the motion of
dislocations are automatically included --- whether they are important
or not.

Unfortunately, this lack of bias in respect to which processes are
important has a major drawback as well: the computational resources
are not focussed on modelling the important processes.  In the average
molecular dynamics simulation a lot of effort is spent on resolving
the thermal vibration of all the atoms.  This leads to restrictions on
the length and time scales that can be modelled, see section
\ref{sec:scales}.

Although no assumptions are made concerning the important processes in
the material under study, molecular dynamics is not completely
independent of theoretical input.  It depends critically on a reliable
model for the interatomic interactions.  The interactions
are described by a potential: a function giving the potential energy
of an atom as a function of the positions of the neighbouring atoms.
The forces on the atoms are then found as the derivatives of the
potential energy.  The quality of the potential is obviously critical
for the quality of the simulation.

The simplest choice for a potential is a pair potential, where the
energy of an atom is given as the sum over the energy of the bonds to
neighbouring atoms:
$E_i = \sum_{j \ne i} E_{\rm bond}(\left|r_j - r_i\right|)$.
% \begin{equation}
%   \label{eq:pairpot}
%   E_i = \sum_{j \ne i} E_{\rm bond}(\left|r_j - r_i\right|).
% \end{equation}
An example of such potentials is the Lennard-Jones potential, which
gives a good description of the interactions between noble gas atoms,
even in the solid phase, but is not suitable for most other condensed
matter systems.

The interactions in a metal or a semiconductor cannot be approximated
by a pair potential, more complicated functional forms are required.
Most simulations of late transition metals use many-body potentials
such as the Embedded Atom Method (EAM) \citep{DaBa84} or the Effective
Medium Theory (EMT) \citep{JaNoPu87,JaStNo96}.  They give a good
description of a group of metals with face-centered cubic crystal
structure, for example the EMT can describe the metals Ni, Cu, Pd, Ag,
Pt and Au, and alloys between them.

In some situations no interatomic potential can describe the
interactions with sufficient accuracy, and one has to use quantum
mechanical calculations where the energy of the system being studied
is found by solving Schr\"odinger's equation for the electrons in the
material.  This is exceedingly time consuming, and limits the size of
the simulation to around a hundred atoms.  Fortunately, many materials
properties do not depend in a sensitive way on the interatomic
interactions, allowing one to use the less computationally expensive
interatomic potentials.  For example, the mechanical behaviour of
nanocrystalline nickel, copper and palladium is experimentally very
similar, and does therefore not depend critically on details in the
interatomic interactions.  One can therefore expect that a reasonable
interatomic potential such as EAM or EMT will capture the essential
physics.  In situations where chemical effects seem to be important,
such as various forms of chemical embrittlement, one has to be more
careful and quantum mechanical methods may be necessary.

The amount of trust that can be placed in the description of the
interatomic interactions should ultimately be determined by comparison
with experiments, but in most cases must be based on the experience
gained from similar simulations.  It is the opinion of this author
that in most simulations of nanocrystalline metals the potential gives
a sufficiently accurate description of the interactions, and that the
error introduced through the potential is probably smaller than errors
introduced by other factors.

\subsection{Limitations on the length and time scales.}
\label{sec:scales}

Since molecular dynamics works by numerically solving Newton's second
law for all the atoms, it captures all the motion of the atoms
including the thermal vibrations.  The time step used when integrating
the equation of motion must therefore be very small.  In typical
simulations the time step is 5 femtoseconds or smaller.  This limits
the simulations to very short times, typically even a few nanoseconds
of simulated time is computationally very expensive, at least for
systems with many atoms.

One consequence of the short time-scale is that most diffusional
processes are excluded from the simulation, as they occur on much
longer time scales.  In connection with nanocrystalline metals it
means for example that diffusional creep cannot be observed in
molecular dynamics simulations.  Fortunately, experimental
measurements of creep rates in nanocrystalline metals indicate that
diffusional creep is not a large effect \citep{NiWeSi91,NiWa91}.
Other thermally activated processes with activation energies near room
temperature will also be suppressed by the short time scale.  In many
cases these processes can be identified for example by simulations at
elevated temperatures, and then studied with more specialized
simulation techniques.

As the time scale of the simulation is very short, very high strain
rates are required to obtain an appreciable deformation during the
simulation --- strain rates around $10^8\,{\rm s}^{-1}$ are typically
used, and one can be worried if this makes the results unreliable.
The strain rate ``gap'' between simulations and experiment should not
be ignored, but it is not an insurmountable obstacle.  Although
stress-strain curves of nanocrystalline metals are strain rate
dependent both in experiments \citep{GrLoCaVaAl97} and simulations
(Fig.~\ref{fig:strainrate}), the change in stresses is typically measured in
percent (or tens of percent) when the strain rate is changed by
several decades.
\begin{Figure}
  \begin{center}
    \epsfig{file=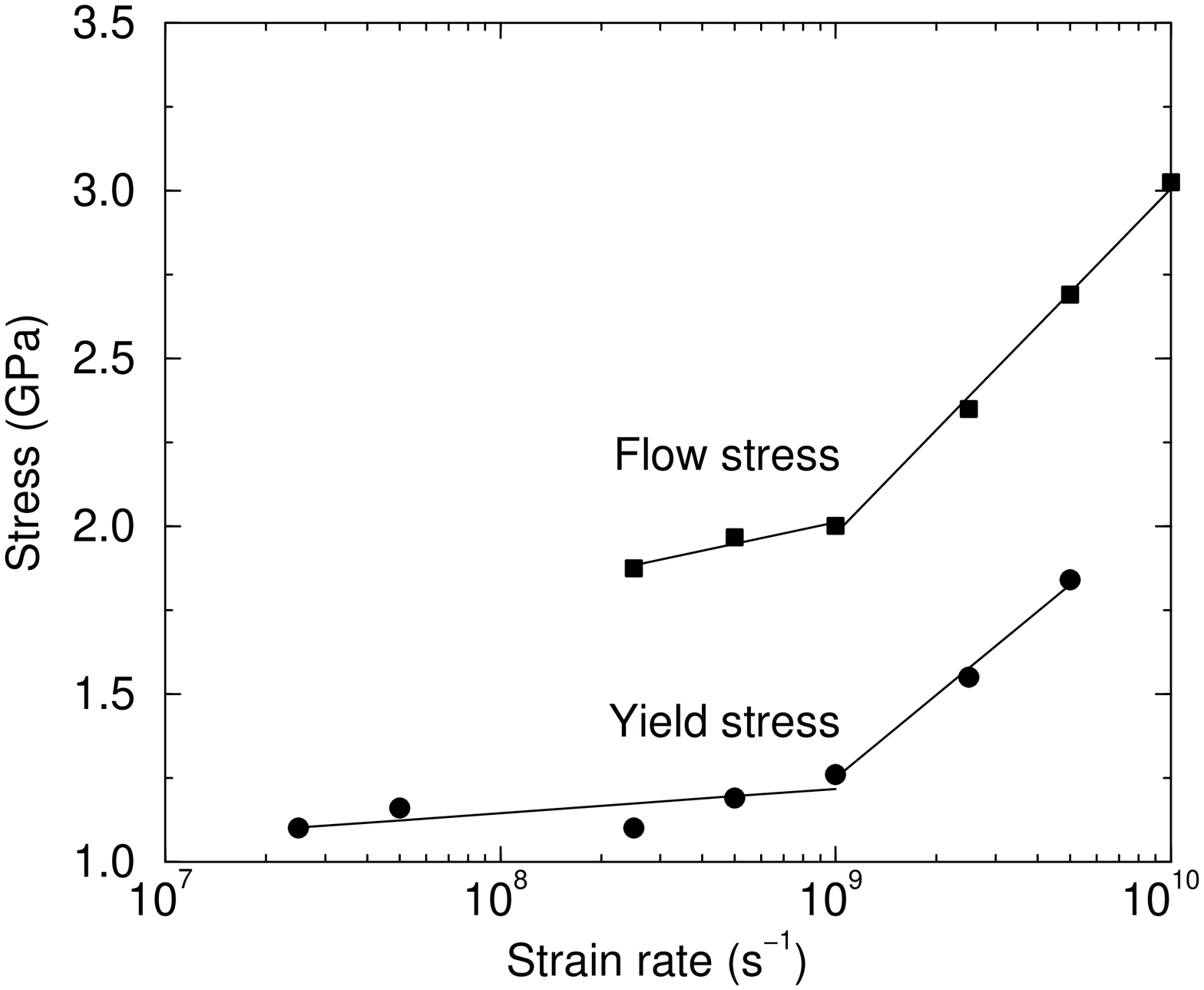, angle=-90, width=0.6\linewidth}
    \caption{The effect of varying the strain rate in simulations of
      nanocrystalline copper with an average grain size of 5.2\,nm.
      The same system was deformed at the same temperature (300\,K)
      but at different strain rates, and the yield and flow stresses
      were extracted from the stress-strain curves.  Both are seen to
      vary with the strain rate ($\dot\varepsilon$), strongest for
      $\dot\varepsilon > 10^{\,9}\, {\rm s}^{-1}$.  Reproduced from
      \citet{ScVeDiJa99}.}
    \label{fig:strainrate}
  \end{center}
\end{Figure}

Since atomic-scale simulations are based on the individual atoms, the
size of the system that can be simulated is limited by the number of
atoms that can be managed in the simulation.  For realistic
simulation, i.e.\ for simulations that cover more than a few
picoseconds and use realistic interatomic potentials, the number of
atoms is limited to a few tens of millions, even using the largest
parallel computers.  This severely limits the grain sizes of the
polycrystalline metal under study.

Several methods have been proposed to overcome or circumvent these
limitations.  In the quasi-continuum method
\citep{TaOrPh96,ShMiTaRoPhOr99} the number of degrees of freedom
is reduced by using ``representative'' atoms to represent volumes of
space where atomic resolution is not necessary.  This allows for
larger systems while maintaining atomic resolution in the relevant
regions.  Unfortunately, it is difficult to use quasi-continuum methods
at finite temperatures \citep[see for example][]{CuCe01}.

A number of methods have also been proposed to deal with the
limitations in time-scale.  They all have in common that a correct
description of the details of the atomic vibrations is sacrificed,
while attempting to move the system between various states with the
correct rates.  Kinetic Monte Carlo \citep{Vo86} and
related methods require that all possible atomic-scale processes are
known a priori, whereas hyperdynamics \citep{Vo97,Vo97b} and
temperature-accelerated dynamics \citep{SoVo00} collect that
information during the simulation.

\subsection{Boundary conditions.}
\label{sec:boundary}

As the number of atoms in a simulation is limited, one is effectively
working with an exceedingly small sample.  To limit the artifacts
caused by the small sample one usually employs periodic boundary
conditions, where the simulation cell is repeated infinitely in all
directions.  In this way no atoms are near free surfaces, and the
system behaves as if it is deep inside the bulk of a material.

In some cases it is relevant to use free boundary conditions in some
directions.  For example, \citet{DeSw01} have simulated a
nanocrystalline film by having free boundary conditions in one
direction, and periodic boundary conditions in the two other.  They
find that the surfaces cause an increase in the dislocation activity
in the sample and a significant increase in the plastic deformation.
The effects of the free surfaces appear to be localized in a layer of
approximately the same thickness as the grain size.  There do not
appear to be \emph{qualitative} changes caused by the free surfaces.
In simulations of ultra-fast deformation of single crystals
\citep{ScLeSi01} film and bulk geometries are also seen to give
qualitatively the same results, whereas wire geometry (with free
boundary conditions in two directions and periodic boundary conditions
in the third) give qualitatively very different results.

\subsection{Initial configuration.}
\label{sec:initialconf}

Just as an atomic-scale simulation provides all details of the atomic
motion during the simulation, it also \emph{requires} all these
details when starting.  It is the responsibility of the user to
provide a complete, atomic-scale description of the initial
configuration of the material.  This can be a major obstacle, as the
full atomic configuration of real-world materials is rarely known, and
can therefore be a major source of error in computer simulations.  A
bad initial configuration (a ``bad sample'') can lead to distorted
results --- a problem simulations share with experiments!

The problem may sound more severe than it is, since in most cases
details of the atomic configuration do not impact the macroscopic
information that one wishes to gather from the simulation.  For
example, the positions of a few atoms may influence exactly when and
where dislocations move through grains, but is unlikely to influence
macroscopic quantities such as yield stress.  Nevertheless, often
simulations should be repeated with different initial configurations
just as experiments are often repeated with different samples.  In
Fig.~\ref{fig:stressCu} the three lower stress-strain curves are the
average over four different simulations with different, randomly
generated grain structures.

Some ingenuity is normally required to generate realistic initial
configurations.  Several methods have been used to generate
nanocrystalline samples. \citet{PhWoGl95} generate a nanocrystalline
sample by a computer simulation where a liquid is solidified in the
presence of crystal nuclei, i.e.\ small spheres of atoms held fixed in
a crystalline lattice.  The system was then quenched a nanocrystalline
metal was formed when the liquid crystallized around the seeds.  The
rapid solidification leads to a large number of stacking faults and
other defects in the grains.  Most other groups have generated the
nanocrystalline samples using a Voronoi construction.  A set of grain
centers are chosen randomly, and the part of space closer to a given
center than to any other center is filled with atoms in a randomly
oriented crystalline lattice.  The sample is then annealed briefly to
remove unfavorable atomic configurations from the grain boundaries,
see \citet{ScVeDiJa99} for details.

The samples generated by these methods are very ``clean'', they do not
contain impurity atoms or large defects like pores and voids.  This is
an advantageous simplification when the simulation is used to gain
understanding of the deformation mechanism.  But it is a disadvantage
when comparing directly to experimental studies, where it is likely
that impurities and defects are dominating the mechanical properties
\citep[see for example][]{AgElYoHeWe98,MoMo97,SaYoWe97}.
Impurities can of course be introduced into computer simulations as
well.  As expected, porosity has a dramatic effect on the mechanical
properties \citep{ScVeDiJa99}.  Studies of impurities are limited by
the lack of interatomic potentials giving a reasonable description of
the embedding of a typical impurity atom (such as oxygen) in a metal.
In a single study, silver has been introduced in the grain boundaries
of copper with little effect \citep{ScVeJa99}.

\subsection{Interpretation.}
\label{sec:interpretation}

Computer simulations rarely produce the desired new knowledge
directly, considerable interpretation is usually required to extract
the science from the simulation.  There are, therefore, all the same
possibilities for mis- and over-interpretations as is present in
experimental work, although the full access to the atomic coordinates
often makes the interpretation process easier and more reliable.

\section{DISCUSSION AND CONCLUSIONS}
\label{sec:conclusions}

Atomic-scale computer simulations can reveal important information
about the deformation mechanism of nanocrystalline metals, information
it would be difficult to obtain from other sources.  Simulations have
shown that the grain boundaries are participating directly in the
deformation process, and if the grain size is sufficiently small the
main deformation mechanism is grain boundary sliding, leading to a
reverse Hall-Petch effect.

It can be difficult to evaluate the trustworthiness and reliability of
computer simulations.  There appears to be a large set of factors that
can influence their reliability in a negative way.  However, computer
simulations are in that respect no different from most other
techniques in materials science --- most techniques, experimental or
theoretical, have their own pitfalls.  As a relative newcomer to the
field, it is natural that atomic-scale computer simulations are
greeted with some healthy scepticism.  The atomic-scale focus of the
technique places severe limitations on what can be done, but it also
makes the approach relatively free from bias from its underlying
models, allowing unexpected phenomena to appear in what is often
called a ``computer experiment''.  If some care it taken when choosing
the method and the subject under study, atomic-scale simulations can
be a reliable source of information which would be difficult or
impossible to obtain in other ways.

\section*{ACKNOWLEDGMENTS}

The author wishes to thank Tejs Vegge who did some of the
simulations.  Center for Atomic-scale Materials Physics is sponsored
by the Danish National Research Council.  The author gratefully
acknowledges financial support from the Danish Research Agency through
grant No.~5020-00-0012.

\section*{REFERENCES}
\setlength{\parskip}{0pt} \renewcommand{\refname}{}
% Fix that volume numbers should be underlined, but makebst did not
% support that, and I used emphasis instead.

% Manually fix titles of series to remove underline/emphasis.
% Manually correct eds to Eds and replace preceding comma with period.
\renewcommand{\emph}{\uline}
%\bibliographystyle{newrisoe}
%\bibliography{js}

\begin{thebibliography}{45}
\expandafter\ifx\csname natexlab\endcsname\relax\def\natexlab#1{#1}\fi
\expandafter\ifx\csname url\endcsname\relax
  \def\url#1{\texttt{#1}}\fi
\expandafter\ifx\csname urlprefix\endcsname\relax\def\urlprefix{URL }\fi

\bibitem[{Agnew et~al.(1998)Agnew, Elliott, Youngdahl, Hemker, and
    Weertman}]{AgElYoHeWe98} Agnew, S.~R., Elliott, B.~R., Youngdahl,
  C.~J., Hemker, K.~J., and Weertman, J.~R. (1998). Structure and
  mechanical properties of nanocrystalline metals --- with
  opportunities for modeling. In: Modelling of Structure and Mechanics
  of Materials from Microscale to Product.  Eds. J.~V. Carstensen et
  al., Proceedings of the 19th Ris{\o} International Symposium on
  Materials Science (Ris{\o} National Laboratory, Roskilde), 1--14.

\bibitem[{Chen(1995)}]{Ch95}
Chen, D. (1995). Structural modeling of nanocrystalline materials. Comput.
  Mater. Sci. \emph{3}, 327--333.

\bibitem[{Curtarolo and Ceder(2001)}]{CuCe01}
Curtarolo, S. and Ceder, G. (2001). Dynamics of a non homogeneously coarse
  grained system. (preprint, http://arXiv.org/abs/cond-mat/0106263).

\bibitem[{Daw and Baskes(1984)}]{DaBa84}
Daw, M.~S. and Baskes, M.~I. (1984). Embedded-atom method: Derivation and
  application to impurities, surfaces, and other defects in metals. Phys. Rev.
  B \emph{29}, 6443--6453.

\bibitem[{Derlet and Van~Swygenhoven(2001)}]{DeSw01}
Derlet, P.~M. and Van~Swygenhoven, H. (2001). The role played by two parallel
  free surfaces in the deformation mechanism of nanocrystalline metals: A
  molecular dynamics simulation. Phil. Mag. A (in press).

\bibitem[{Faken and J\'onsson(1994)}]{FaJo94}
Faken, D. and J\'onsson, H. (1994). Systematic analysis of local atomic
  structure combined with 3d computer graphics. Comput. Mater. Sci. \emph{2},
  279--286.

\bibitem[{Fitzsimmons et~al.(1991)Fitzsimmons, Eastman, M\"uller-Stach, and
  Wallner}]{FiEaMuWa91}
Fitzsimmons, M.~R., Eastman, J.~A., M\"uller-Stach, M., and Wallner, G. (1991).
  Structural characterization of nanometer-sized crystalline {Pd} by
  x-ray-diffraction techniques. Phys. Rev. B \emph{44}, 2452--2460.

\bibitem[{Gray et~al.(1997)Gray, Lowe, Cady, Valiev, and
  Aleksandrov}]{GrLoCaVaAl97}
Gray, III, G.~T., Lowe, T.~C., Cady, C.~M., Valiev, R.~Z., and Aleksandrov,
  I.~V. (1997). Influence of strain rate \& temperature on the mechanical
  response of ultrafine-grained {Cu}, {Ni} and {Al-4Cu-0.5Zr}. NanoStruct.
  Mater. \emph{9}, 477--480.

\bibitem[{Hall(1951)}]{Ha51}
Hall, E.~O. (1951). The deformation and ageing of mild steel: {III}
  {D}iscussion of results. Proc. Phys. Soc. London \emph{B64}, 747--753.

\bibitem[{Heino and Ristolainen(2001)}]{HeRi01}
Heino, P. and Ristolainen, E. (2001). Strength of nanoscale polycrystalline
  copper under shear. Phil. Mag. A \emph{81}, 957--970.

\bibitem[{Honeycutt and Andersen(1987)}]{HoAn87}
Honeycutt, J.~D. and Andersen, H.~C. (1987). Molecular dynamics study of
  melting and freezing of small lennard-jones clusters. J. Phys. Chem.
  \emph{91}, 4950--4963.

\bibitem[{Jacobsen et~al.(1987)Jacobsen, N{\o}rskov, and Puska}]{JaNoPu87}
Jacobsen, K.~W., N{\o}rskov, J.~K., and Puska, M.~J. (1987). Interatomic
  interactions in the effective-medium theory. Phys. Rev. B \emph{35},
  7423--7442.

\bibitem[{Jacobsen et~al.(1996)Jacobsen, Stoltze, and N{\o}rskov}]{JaStNo96}
Jacobsen, K.~W., Stoltze, P., and N{\o}rskov, J.~K. (1996). A semi-empirical
  effective medium theory for metals and alloys. Surf. Sci. \emph{366},
  394--402.

\bibitem[{Keblinski et~al.(1997)Keblinski, Phillpot, Wolf, and
  Gleiter}]{KePhWoGl97}
Keblinski, P., Phillpot, R., Wolf, D., and Gleiter, H. (1997). Amorphous
  structure of grain boundaries and grain junctions in nanocrystalline silicon
  by molecular-dynamics simulation. Acta Mater. \emph{45}, 987--998.

\bibitem[{Keblinski et~al.(1999)Keblinski, Wolf, Phillpot, and
  Gleiter}]{KeWoPhGl99}
Keblinski, P., Wolf, D., Phillpot, S.~R., and Gleiter, H. (1999). Structure of
  grain boundaries in nanocrystalline palladium by molecular dynamics
  simulation. Scripta Mater. \emph{41}, 631--636.

\bibitem[{Morris(1998)}]{Mo98}
Morris, D.~G. (1998). Mechanical Behavior of Nanostructured Materials, vol.~2
  of Materials Science Foundations (Trans Tech Pub., Uetikon-Z\"urich).

\bibitem[{Morris and Morris(1997)}]{MoMo97}
Morris, D.~G. and Morris, M.~A. (1997). Hardness, strength, ductility and
  toughness of nanocrystalline materials. Mater. Sci. Forum \emph{235-238},
  861--872.

\bibitem[{Nieh and Wadsworth(1991)}]{NiWa91}
Nieh, T.~G. and Wadsworth, J. (1991). {H}all-{P}etch relation in
  nanocrystalline solids. Scripta Met. Mater. \emph{25}, 955--958.

\bibitem[{Nieman et~al.(1991)Nieman, Weertman, and Siegel}]{NiWeSi91}
Nieman, G.~W., Weertman, J.~R., and Siegel, R.~W. (1991). Mechanical behavior
  of nanocrystalline {Cu} and {Pd}. J. Mater. Res. \emph{6}, 1012--1027.

\bibitem[{Petch(1953)}]{Pe53}
Petch, N.~J. (1953). The cleavage strength of polycrystals. J. Iron Steel Inst.
  \emph{174}, 25.

\bibitem[{Phillpot et~al.(1995{\natexlab{a}})Phillpot, Wolf, and
  Gleiter}]{PhWoGl95}
Phillpot, S.~R., Wolf, D., and Gleiter, H. (1995{\natexlab{a}}).
  Molecular-dynamics study of the synthesis and characterization of a fully
  dense, three-dimenstional nanocrystalline material. J. Appl. Phys. \emph{78},
  847--860.

\bibitem[{Phillpot et~al.(1995{\natexlab{b}})Phillpot, Wolf, and
  Gleiter}]{PhWoGl95b}
Phillpot, S.~R., Wolf, D., and Gleiter, H. (1995{\natexlab{b}}). A structural
  model for grain boundaries in nanocrystalline materials. Scripta Met. Mater.
  \emph{33}, 1245--1251.

\bibitem[{Rasmussen et~al.(1997)Rasmussen, Jacobsen, Leffers, Pedersen,
  Srinivasan, and J\'onsson}]{RaJaLePeSrJo97}
Rasmussen, T., Jacobsen, K.~W., Leffers, T., Pedersen, O.~B., Srinivasan,
  S.~G., and J\'onsson, H. (1997). Atomistic determination of cross-slip
  pathway and energetics. Phys. Rev. Lett. \emph{79}, 3676--3679.

\bibitem[{Sanders et~al.(1997)Sanders, Youngdahl, and Weertman}]{SaYoWe97}
Sanders, P.~G., Youngdahl, C.~J., and Weertman, J.~R. (1997). The strength of
  nanocrystalline metals with and without flaws. Mater. Sci. Eng. A
  \emph{234-236}, 77--82.

\bibitem[{Schi{\o}tz et~al.(1998{\natexlab{a}})Schi{\o}tz, Di~Tolla, and
  Jacobsen}]{ScDiJa98}
Schi{\o}tz, J., Di~Tolla, F.~D., and Jacobsen, K.~W. (1998{\natexlab{a}}).
  Softening of nanocrystalline metals at very small grain sizes. Nature
  \emph{391}, 561--563.

\bibitem[{Schi{\o}tz et~al.(2001)Schi{\o}tz, Leffers, and Singh}]{ScLeSi01}
Schi{\o}tz, J., Leffers, T., and Singh, B.~N. (2001). Dislocation nucleation
  and vacancy formation during high-speed deformation of fcc metals. Phil. Mag.
  Lett. \emph{81}, 301--309.

\bibitem[{Schi{\o}tz et~al.(1998{\natexlab{b}})Schi{\o}tz, Vegge, Di~Tolla, and
  Jacobsen}]{ScVeDiJa98}
Schi{\o}tz, J., Vegge, T., Di~Tolla, F.~D., and Jacobsen, K.~W.
  (1998{\natexlab{b}}). Simulations of mechanics and structure of nanomaterials
  --- from nanoscale to coarser scales. In: Modelling of Structure and
  Mechanics of Materials from Microscale to Product.
  Eds. J.~V. Carstensen et al.,
  Proceedings of the 19th Ris{\o} International Symposium on Materials Science
  (Ris{\o} National Laboratory, Roskilde), 133--148.

\bibitem[{Schi{\o}tz et~al.(1999{\natexlab{a}})Schi{\o}tz, Vegge, Di~Tolla, and
  Jacobsen}]{ScVeDiJa99}
Schi{\o}tz, J., Vegge, T., Di~Tolla, F.~D., and Jacobsen, K.~W.
  (1999{\natexlab{a}}). Atomic-scale simulations of the mechanical deformation
  of nanocrystalline metals. Phys. Rev. B \emph{60}, 11971--11983.

\bibitem[{Schi{\o}tz et~al.(1999{\natexlab{b}})Schi{\o}tz, Vegge, and
  Jacobsen}]{ScVeJa99}
Schi{\o}tz, J., Vegge, T., and Jacobsen, K.~W. (1999{\natexlab{b}}).
  Atomic-scale modeling of the deformation of nanocrystalline metals. In:
  Multiscale Modeling of Materials.  Eds. V.~V. Bulatov et~al., vol. 538 of
  Mater. Res. Soc. Symp. Proc., 299--308.

\bibitem[{Shenoy et~al.(1999)Shenoy, Miller, Tadmor, Rodney, Phillips, and
  Ortiz}]{ShMiTaRoPhOr99}
Shenoy, V.~B., Miller, R., Tadmor, E.~B., Rodney, D., Phillips, R., and Ortiz,
  M. (1999). An adaptive finite element approach to atomic-scale mechanics ---
  the quasicontinuum method. J. Mech. Phys. Solids \emph{47}, 611--642.

\bibitem[{S{\o}rensen and Voter(2000)}]{SoVo00}
S{\o}rensen, M.~R. and Voter, A.~F. (2000). Temperature-accelerated dynamics
  for simulation of infrequent events. J. Chem. Phys. \emph{112}, 9599--9605.

\bibitem[{Stern et~al.(1995)Stern, Siegel, Newville, Sanders, and
  Haskel}]{StSiNeSaHa95}
Stern, E.~A., Siegel, R.~W., Newville, M., Sanders, P.~G., and Haskel, D.
  (1995). Are nanophase grain boundaries anomalous? Phys. Rev. Lett. \emph{75},
  3874--3877.

\bibitem[{Tadmor et~al.(1996)Tadmor, Ortiz, and Phillips}]{TaOrPh96}
Tadmor, E.~B., Ortiz, M., and Phillips, R. (1996). Quasicontinuum analysis of
  defects in solids. Phil. Mag. A \emph{73}, 1529--1564.

\bibitem[{Van~Swygenhoven and Caro(1997)}]{SwCa97}
Van~Swygenhoven, H. and Caro, A. (1997). Plastic behavior of nanophase {Ni}: a
  molecular dynamics computer simulation. Appl. Phys. Lett. \emph{71},
  1652--1654.

\bibitem[{Van~Swygenhoven and Caro(1998)}]{SwCa98}
Van~Swygenhoven, H. and Caro, A. (1998). Plastic behavior of nanophase metals
  studied by molecular dynamics. Phys. Rev. B \emph{58}, 11246--11251.

\bibitem[{Van~Swygenhoven et~al.(2000)Van~Swygenhoven, Farkas, and
  Caro}]{SwFaCa00}
Van~Swygenhoven, H., Farkas, D., and Caro, A. (2000). Grain-boundary structures
  in polycrystalline metals at the nanoscale. Phys. Rev. B \emph{62}, 831--838.

\bibitem[{Van~Swygenhoven et~al.(1999)Van~Swygenhoven, Spaczer, Caro, and
  Farkas}]{SwSpCaFa99}
Van~Swygenhoven, H., Spaczer, M., Caro, A., and Farkas, D. (1999). Competing
  plastic deformation mechanisms in nanophase metals. Phys. Rev. B \emph{60},
  22--25.

\bibitem[{Vegge et~al.(2000)Vegge, Rasmussen, Leffers, Pedersen, and
  Jacobsen}]{VeRaLePeJa00}
Vegge, T., Rasmussen, T., Leffers, T., Pedersen, O.~B., and Jacobsen, K.~W.
  (2000). Determination of the of rate cross slip of screw dislocations. Phys.
  Rev. Lett. \emph{85}, 3866--3869.

\bibitem[{Vegge et~al.(2001)Vegge, Rasmussen, Leffers, Pedersen, and
  Jacobsen}]{VeRaLePeJa01}
Vegge, T., Rasmussen, T., Leffers, T., Pedersen, O.~B., and Jacobsen, K.~W.
  (2001). Atomistic simulations of cross-slip of jogged screw dislocations in
  copper. Phil. Mag. Lett. \emph{81}, 137--144.

\bibitem[{Voter(1986)}]{Vo86}
Voter, A.~F. (1986). Classically exact overlayer dynamics: diffusion of rhodium
  clusters on {Rh(100)}. Phys. Rev. B \emph{34}, 6819--29.

\bibitem[{Voter(1997{\natexlab{a}})}]{Vo97}
Voter, A.~F. (1997{\natexlab{a}}). Hyperdynamics: Accelerated molecular
  dynamics of infrequent events. Phys. Rev. Lett. \emph{78}, 3908--3911.

\bibitem[{Voter(1997{\natexlab{b}})}]{Vo97b}
Voter, A.~F. (1997{\natexlab{b}}). A method for accelerating the molecular
  dynamics simulation of infrequent events. J. Chem. Phys. \emph{106},
  4665--4677.

\bibitem[{Yip(1998)}]{Yi98}
Yip, S. (1998). The strongest size. Nature \emph{391}, 532--533, (N\&V).

\bibitem[{Zhu and Averback(1996)}]{ZhAv96}
Zhu, H. and Averback, R.~S. (1996). Sintering of nano-particle powders:
  simulations and experiments. Materials and Manufacturing Processes \emph{11},
  905--923.

\bibitem[{Zhu et~al.(1987)Zhu, Birringer, Herr, and Gleiter}]{ZhBiHeGl87}
Zhu, X., Birringer, R., Herr, U., and Gleiter, H. (1987). X-ray diffraction
  studies of nanometer-sized crystalline materials. Phys. Rev. B \emph{35},
  9085--9090.

\end{thebibliography}

\end{document}